\documentclass[referee]{aa} 
\usepackage{graphicx}
\usepackage{natbib}
\bibpunct{(}{)}{;}{a}{}{,}
\defcitealias{ca89}{CCM}
\begin{document}
   \title{Optical observations of a newly identified compact galaxy
          group \\ near the Zone of Avoidance\thanks{Based on data obtained 
	  at Calar Alto (Spain) and Asiago (Italy) observatories.}}


   \author{S. Temporin
          \inst{1}
          \and
          S. Ciroi\inst{2}
          }


   \institute{Institut f\"ur Astrophysik, Universit\"at Innsbruck,\\
              Technikerstra\ss e 25, A-6020 Innsbruck, Austria\\
              \email{giovanna.temporin@uibk.ac.at}
         \and
             Dipartimento di Astronomia dell' Universit\'a di Padova,\\ 
	     Vicolo dell'Osservatorio 2, I-35122 Padova, Italy\\
             \email{ciroi@pd.astro.it}
             }

   \date{Received ??; accepted ??}

   \abstract{
   We have identified a new group of galaxies, CG~J0247+44.9, at low galactic 
   latitude ($l=143^{\circ}.64$, $b=-13^{\circ}.29$), which satisfies 
   Hickson's criteria \citep{hick97} for Compact Groups (CGs). 
   Our group consists of six members, two of which are in close 
   interaction (IRAS 02443+4437). 
   
   We present here optical photometry
   (BVRI) and low resolution spectroscopy of the individual galaxies
   and investigate the global properties of the group. 
   Our morphological analysis reveals that two out of the six objects are 
   lenticular galaxies. The others are spirals showing emission lines 
   in their spectra through which we could classify them as a starburst 
   galaxy (the spiral member of the IRAS 02443+4437 close pair), 
   a Seyfert 2, a LINER and a weak HII galaxy. 
   Since the S0/Sa is the prevailing morphology for the galaxies of this group, 
   which is also characterized by a short crossing time and a relatively high 
   velocity dispersion, we suggest that CG~J0247+44.9 is a dynamically old compact group.

   }

   \titlerunning{The Compact Group CG~J0247+44.9}
   \maketitle
%

\section{Introduction}

   As it is well known, our view of the extragalactic space at low Galactic latitudes is 
   obscured by a patchy layer of dust, 
   whose inner part ($|b| \la 10$\degr) is called Zone of Avoidance (ZoA).
   However, numerous multi-wavelength surveys carried out by different research 
   groups \citep[e.g.][and references therein]{nt97,kk00,wb00} in the last decades
   led to the detection of tens of thousand of galaxies in this band of the
   Milky Way.
   Such a huge galaxy database has a high potential both for extragalactic
   studies and for studies of our own Galaxy.
   In particular we started a project (S. Temporin et~al., in preparation) devoted to 
   investigate the dust distribution in the Milky Way in dependence on
   Galactic longitude and latitude. This purpose will be achieved by evaluating 
   the total foreground
   Galactic extinction along many lines-of-sight to galaxies in the 
   ZoA and its vicinity through BVRI CCD-photometry.
   Indeed, depending on their morphological type, galaxies have 
   well-defined intrinsic total and effective optical colors \citep{bw95}, 
   which can be used to determine the color 
   excess and then the visual absorption A$\rm _V$. 
   In order to largely avoid the fine-scale structure of Galactic dust 
   (clumpiness) we selected close pairs and/or multiple systems of galaxies 
   to reduce the errors in the determination of Galactic 
   extinction along a specific line-of-sight. 
   
   As an aside we have performed a spectroscopic follow-up of these selected 
   systems, whose physical properties are still widely unknown, in order to 
   establish, through the determination of their radial velocities, which of 
   them are physically bound and not only chance projections 
   on the plane of the sky.
   We have identified in this way a new compact galaxy group at low galactic 
   latitude ($l=143^{\circ}.64$, $b=-13^{\circ}.29$), to which we 
   refer here as CG~J0247+44.9, containing six members. Two of them form a close 
   interacting pair, already known as IRAS~02443+4437, other three galaxies 
   are cataloged in the 2MASS survey \citep{huc00} and the last one is 
   uncataloged.   
   We present here the analysis of photometric (Sect. 2) and spectroscopic data (Sect. 3)
   of this group. We discuss the individual properties of its 
   members in Sect. 4 and the global properties of the group in Sect. 5.
   The main results are summarized in Sect. 6.
   

\section{BVRI photometry}

Data in the broad-band BVRI Johnson-Cousins system were obtained 
under photometric conditions at the
1.23 m telescope of the Calar Alto Observatory (Spain) in January 1999.
The images covered  a field of view of
$\sim$ 8$^{\prime}$.5$\times$8$^{\prime}$.5 with a spatial scale of 0.5 arcsec pixel$^{-1}$.
Observational details are summarized in Table~\ref{log}.

  \begin{table}
      \caption[]{Observation Logs}
         \label{log}
     $$
         \begin{tabular}{cccccccc}
            \hline
            \noalign{\smallskip}
            Telescope & Instrument & Filter/Grism &  Date & T$_{\rm exp}$ & Seeing & 
	    Scale & Slit \\
	    & & & & (s) & ('') & (''/px) & ('') \\
            \noalign{\smallskip}
            \hline
            \noalign{\smallskip}
            Calar Alto 1.23m & CCD-camera & Johnson B & 1999-01-06 & 2100 & 1.4 & 0.5 & ... \\
            Calar Alto 1.23m & CCD-camera & Johnson V & 1999-01-06 & 1200 & 1.1 & 0.5 & ... \\
            Calar Alto 1.23m & CCD-camera & Cousins R & 1999-01-06 &  900 & 1.0 & 0.5 & ... \\
            Calar Alto 1.23m & CCD-camera & Cousins I & 1999-01-06 &  300 & 1.1 & 0.5 & ... \\
	    Asiago 1.82m & AFOSC & \#4 & 2001-09-13 & 1800 & 1.2 & 0.47 & 1.26 \\
	    Asiago 1.82m & AFOSC & \#4 & 2001-11-25 & 1800 & 2.2 & 0.47 & 2.10 \\
	    Asiago 1.82m & AFOSC & \#4 & 2001-11-26 & 2x1800 & 2.2 & 0.47 & 2.10 \\	    
	    \noalign{\smallskip}
            \hline
         \end{tabular}
     $$
   \end{table}

The images were reduced with IRAF\footnote{IRAF is distributed by the National Optical 
Astronomy Observatories, which are operated by the Association of Universities
for Research in Astronomy, Inc., under cooperative agreement with the 
National Science Foundation.} following the standard steps of bias 
subtraction, flat fielding and cosmic rays removal. An additional defringing
procedure was applied to the I band image. 
A \citet{la92} standard field was observed three times during the night for
photometric calibration purposes.
The calibration constants (see Table~\ref{calib}) were determined by means of the 
IRAF package PHOTCAL
through a uniformly weighted fit of the following transformation equations:

\begin{equation}
 B = b + B_0 - k_{\rm B} X{\rm _B} + C_{\rm B} (B-V)
\label{Btransf}
\end{equation}
\begin{equation}
 V = v + V_0 - k_{\rm V} X{\rm _V} + C_{\rm V} (B-V)
\label{Vtransf}
\end{equation}
\begin{equation}
R = r + R_0 - k_{\rm R} X{\rm _R} + C_{\rm R} (V-R)
\label{Rtransf}
\end{equation}
\begin{equation}
I = i + I_0 - k_{\rm I} X_{\rm I} + C_{\rm I} (R-I)
\end{equation}

where $B$, $V$, $R$, and $I$ are calibrated magnitudes, $b$, $v$, $r$, and $i$ are
sky-subtracted instrumental magnitudes normalized to 1 s exposure and 
$X{\rm _B}$, $X{\rm _V}$, $X{\rm _R}$, and $X{\rm _I}$ are the airmasses at the time of 
the observations.

   \begin{table}
      \caption[]{Photometric Calibration Constants$^{\mathrm{a}}$.}
         \label{calib}
     $$
         \begin{tabular}{cccc}
            \hline
            \noalign{\smallskip}
            Band & $\rm M_0$ & $k_{\rm M}$ &  $C_{\rm M}$ \\
            \noalign{\smallskip}
            \hline
            \noalign{\smallskip}
            B & 22.193 $\pm$ 0.054 & 0.227 $\pm$ 0.041 & 0.110 $\pm$ 0.009 \\
	    V & 22.624 $\pm$ 0.041 & 0.138 $\pm$ 0.030 & -0.028 $\pm$ 0.007 \\
	    R & 23.102 $\pm$ 0.036 & 0.086 $\pm$ 0.026 & 0.064 $\pm$ 0.009 \\
	    I & 22.526 $\pm$ 0.162 & 0.071 $\pm$ 0.117 & 0.274 $\pm$ 0.061 \\
            \noalign{\smallskip}
            \hline
         \end{tabular}
     $$
     \begin{list}{}{}
\item[$^{\mathrm{a}}$] With $\rm M_0$, $k_{\rm M}$, and $C_{\rm M}$ we indicate
the zero point, the extinction coefficient and color term, respectively, in each
of the four photometric bands.
\end{list}
   \end{table}
   
 We have identified seven galaxies in the field, two of which form an interactive pair
 known as IRAS 02443+4437 (Fig.~\ref{ident}). 
 We have studied the morphology of each galaxy and calculated the total magnitudes in the
 four bands by means of GIM2D \citep{sim98,sim01}, an IRAF package written to perform an 
 automated bulge-disk decomposition of the surface brightness profile. 
 GIM2D is well suited for low signal-to-noise images of
distant galaxies since it has the considerable advantage of including a PSF deconvolution 
in the bi-dimensional fit.
The effects of the distance in dimming the outer parts of galaxies' disks can be 
considered comparable to the effects produced by heavy Galactic foreground extinction in nearby galaxies.
Therefore the suitability of the program to the bulge-disk decomposition of distant galaxies makes it
also particularly appropriate in the case of low-Galactic-latitude galaxies -- like
those studied in this work -- which might be heavily affected by foreground Galactic
extinction.

The program gives as output, the values of the best-fit parameters
and asymmetry indices, besides an image of the galaxy model and a residual image obtained 
by subtracting the model from the original one. 
This fitting method has already been successfully applied
by \citet{tr01} to a sample of galaxies belonging to poor galaxy groups 
and spanning a wide range in morphological type. They found that the ratio
between the bulge and the total luminosity (B/T) can be used on average as a robust 
morphology indicator to discriminate between early-type bulge-dominated (B/T $>$ 0.4) 
and late-type disk-dominated (B/T $<$ 0.4) galaxies.

The PSF images in the four photometric bands to be used during the galaxy fitting procedure
were obtained with the DAOPHOT \citep{sh88} package inside IRAF.
The isophotal area of the galaxies was defined 
by means of the galaxy photometry package SExtractor V2.2.2 \citep{ba96}.
The fit uses an exponential law for the disk, while a classical de Vaucouleurs $r^{1/4}$-law 
\citep{dv48} or
a S\'ersic profile \citep{se68} can be chosen for the bulge. In the last case, the
$n$ index of the S\'ersic law is one of the fitting parameters.
We attempted fits of the seven galaxies with a S\'ersic law plus exponential disk.
When the $n$ index was equal or very close to 4, we applied a de Vaucouleurs law for the bulge.
In Table~\ref{gim2d} we list the best-fit parameters from GIM2D: bulge to total light ratio (B/T), 
bulge effective radius ($R_e$), ellipticity ($e$), disk scale length ($R_d$), inclination ($i$), 
S\'ersic law index ($n$), half-light radius ($R_{half}$), asymmetry index ($R_A$), together with 
the reduced $\chi^2$. We show in Fig.~\ref{BVRI} the BVRI thumbnails of each 
galaxy and in Fig.\ref{BVRI_r} the corresponding residuals after subtraction of the best-fit
galaxy models.

   \begin{table}
   \centering
      \caption[]{GIM2D best-fit parameters.}
         \label{gim2d}
     $$
         \begin{tabular}{llllllllll}
            \hline
            \noalign{\smallskip}
            Band & B/T &  $R_e~(^{\prime\prime})$ & $e$ & $R_d~(^{\prime\prime})$ & $i~(^{\circ})$
	    & $n$ & $\chi^2$ & $R_{half}~(^{\prime\prime})$ & $R_A$ \\
            \noalign{\smallskip}
            \hline
	    \noalign{\smallskip}
	    \multicolumn{10}{c}{\object{CG~J0247+449a}}	\\    
            \noalign{\smallskip}
	    \hline
            \noalign{\smallskip}
            B & 0.31  & 0.62  & 0.21  & 2.76 & 54 & 0.91 & 1.01 & 2.86 & 0.02\\
	    V & 0.36  & 0.67  & 0.20  & 3.19 & 56 & 1.15 & 1.02 & 2.81 & 0.01\\
	    R & 0.38  & 0.57  & 0.21  & 3.19 & 57 & 1.31 & 1.07 & 2.65 & 0.01 \\
	    I & 0.45  & 0.63  & 0.22  & 3.87 & 58 & 2.33 & 1.00 & 2.70 & 0.00 \\
            \noalign{\smallskip}
            \hline
	    \noalign{\smallskip}
	    \multicolumn{10}{c}{CG~J0247+449b}	\\    
            \noalign{\smallskip}
	    \hline
            \noalign{\smallskip}
	    B & 0.59 & 1.08 & 0.20 & 4.06 & 56 & 4 & 1.05 & 2.74 & 0.02 \\
	    V & 0.57 & 0.97 & 0.28 & 3.82 & 52 & 4 & 1.04 & 2.61 & 0.01 \\
	    R & 0.59 & 0.95 & 0.22 & 4.19 & 53 & 4 & 1.11 & 2.51 & 0.02 \\
	    I & 0.55 & 0.80 & 0.26 & 4.05 & 48 & 4 & 1.03 & 2.57 & 0.01 \\
            \noalign{\smallskip}
            \hline
	    \noalign{\smallskip}
	    \multicolumn{10}{c}{CG~J0247+449c}	\\    
            \noalign{\smallskip}
	    \hline
            \noalign{\smallskip}
	    B & 0.33 & 1.22 & 0.39 & 4.86 & 45 & 4 & 1.00 & 5.57 & 0.00 \\
	    V & 0.42 & 1.65 & 0.33 & 5.12 & 46 & 4 & 1.05 & 5.33 & 0.01 \\
	    R & 0.40 & 1.40 & 0.32 & 5.67 & 46 & 4 & 1.09 & 5.69 & 0.01 \\
	    I & 0.39 & 1.05 & 0.32 & 4.77 & 48 & 4 & 1.00 & 4.78 & 0.01 \\
            \noalign{\smallskip}
            \hline
	    \noalign{\smallskip}
	    \multicolumn{10}{c}{CG~J0247+449d}	\\    
            \noalign{\smallskip}
	    \hline
            \noalign{\smallskip}
	    B & 0.22 & 3.99  & 0.69 & 9.29  & 76 & 1.64 & 1.07 & 12.44 & 0.10 \\
	    V & 0.84 & 14.91 & 0.70 & 34.54 & 85 & 2.48 & 1.22 & 18.63 & 0.07 \\
	    R & 0.60 & 8.88  & 0.70 & 15.29 & 73 & 2.01 & 1.13 & 14.09 & 0.07 \\
	    I & 0.94 & 13.25 & 0.41 & 2.33  & 82 & 1.64 & 1.07 & 12.31 & 0.08 \\
            \noalign{\smallskip}
            \hline
	    \noalign{\smallskip}
	    \multicolumn{10}{c}{CG~J0247+449e}	\\    
            \noalign{\smallskip}
	    \hline
            \noalign{\smallskip}
	    B & 0.32 & 2.87 & 0.64 & 6.06 & 55 & 0.38 & 1.60 & 6.24 & 0.17 \\
	    V & 0.42 & 3.38 & 0.65 & 4.29 & 55 & 0.48 & 2.90 & 4.86 & 0.17 \\
	    R & 0.35 & 2.93 & 0.69 & 4.67 & 55 & 0.41 & 3.00 & 5.06 & 0.16 \\
	    I & 0.39 & 2.86 & 0.67 & 6.30 & 60 & 0.73 & 1.40 & 5.95 & 0.09 \\
            \noalign{\smallskip}
            \hline
	    \noalign{\smallskip}
	    \multicolumn{10}{c}{CG~J0247+449f}	\\    
            \noalign{\smallskip}
	    \hline
            \noalign{\smallskip}
	    B & 0.77 & 3.17 & 0.38 & 4.80 & 60 & 4 & 1.17 & 4.27 & 0.02 \\
	    V & 0.65 & 2.86 & 0.39 & 5.68 & 43 & 4 & 1.16 & 4.98 & 0.04 \\
	    R & 0.73 & 3.04 & 0.42 & 5.92 & 54 & 4 & 1.44 & 4.61 & 0.04 \\
	    I & 0.67 & 2.53 & 0.46 & 4.97 & 45 & 4 & 1.17 & 4.21 & 0.02 \\
            \noalign{\smallskip}
            \hline
	    \noalign{\smallskip}
	    \multicolumn{10}{c}{CG~J0247+449g}	\\    
            \noalign{\smallskip}
	    \hline
            \noalign{\smallskip}
	    B & 0.66 & 2.60 & 0.34 & 3.00 & 69 & 4 & 1.00 & 3.57 & 0.01 \\
	    V & 0.72 & 3.27 & 0.41 & 5.64 & 59 & 4 & 0.99 & 4.86 & 0.01 \\
	    R & 0.64 & 2.12 & 0.40 & 3.23 & 25 & 4 & 1.00 & 3.34 & 0.01 \\
	    I & 0.53 & 1.51 & 0.51 & 3.42 & 38 & 4 & 1.08 & 3.34 & 0.02 \\
            \noalign{\smallskip}
	    \hline
         \end{tabular}
     $$
   \end{table}

The two parameters B/T and $n$ were used for the morphological classification of the galaxies
(see Sect. 4).
Our B/T values in the B-band were compared with those plotted as a function of the morphological
type T by \citet{sdv86}. Additionally we took into account the growing evidence that 
late-type spirals host non-classical bulges, better reproduced by an exponential law ($n$ = 1)
\citep{as94,a98,ca01}.

The inclination correction terms A$_{\rm B_i}$, E(B$-$V)$_{\rm i}$ and the selective extinction
R$_{\rm V}$ have been calculated 
as a function of the morphological type T and of the B-band disk inclination $i$ (instead of the  
ratio R$_{25}$ between the major and minor diameters at the
$\mu_{\rm B}$ = 25 mag arcsec$^{-2}$ isophote) following the
instructions given in the introduction to the Third Reference
Catalogue of Bright Galaxies \citep[RC3]{dv91}. The inclination corrections in the other photometric 
bands were derived by application of the \citep[CCM]{ca89} extinction law.
The inclination corrected total and effective colors B$-$V, V$-$R, and V$-$I of three galaxies,
selected for being sufficiently separated one another and not too much inclined, were used to 
estimate the total foreground Galactic extinction. For this purpose the three colors available
for each galaxy were compared with the
mean values for the relevant morphological type published by \citet{bw95}, and the corresponding 
color excesses E(B$-$V) were calculated and averaged. Finally the weighted mean of the values
obtained for these three objects, E(B$-$V) = 0.29 $\pm$ 0.06 (A$_{\rm B}$ = 1.19 mag, assuming
a selective extinction R$_{\rm V}$ = 3.1) was used as a measure of the Galactic 
extinction in the direction of the galaxy group.
We stress that this value is significantly different from the ones available from 
NED\footnote{NASA Extragalactic Database}, A$_{\rm B}$ = 0.610 and A$_{\rm B}$ = 0.642 mag,
based on the maps of \citet{sch98} and \citet{bh82}, respectively.
This discrepancy could be due to small-scale fluctuations in the foreground extinction
as a consequence of the patchy dust distribution at low Galactic latitudes. 
Such fluctuations cannot be detected by interpolating data from large scale maps like
those cited above.
In Table~\ref{mag} we list observed magnitudes and colors of the seven galaxies in the field
and the corresponding values (in bold) corrected for inclination and Galactic extinction.

   \begin{table}
      \caption[]{Magnitudes and Colors$^{\mathrm{a}}$.}
         \label{mag}
     $$
         \begin{tabular}{llllllll}
            \hline
            \noalign{\smallskip}
            Obj. Id.$^{\mathrm{b}}$ & B &  V & R & I & B$-$V & V$-$R & V$-$I \\
            \noalign{\smallskip}
            \hline
a & 17.96 & 16.75 & 16.07 & 15.38 & 1.21 &  0.68 &  1.38 \\
              & {\bf 16.25} & {\bf 15.47} & {\bf 14.99} & {\bf 14.59} & {\bf 0.78} & {\bf  0.47} & {\bf  0.89} \\
b & 17.10 & 15.95 & 15.24 & 14.55 & 1.15 &  0.71 & 1.40 \\
              &{\bf 15.91} & {\bf 15.05} & {\bf 14.48} &{\bf 13.98} & {\bf 0.86}& {\bf 0.58}& {\bf 1.07} \\
c & 16.64  &15.41 & 14.62 & 14.00 & 1.23  & 0.79 & 1.41 \\	      
              &{\bf 15.14} & {\bf 14.28} &  {\bf 13.67}& {\bf 13.29}  & {\bf 0.86} & {\bf 0.61} & {\bf 0.99} \\
d &18.12 & 16.90 & 16.51 & 15.70  &1.23  & 0.39& 1.20 \\
              &{\bf 15.11 }& {\bf 14.58}& {\bf 14.54} &{\bf 14.22 }& {\bf 0.54}& {\bf 0.04} & {\bf 0.36} \\
e &16.13  &15.69 & 14.98 & 14.30 & 0.44 &  0.72 & 1.39 \\
              &{\bf 14.22} & {\bf 14.23} & {\bf 13.74} & {\bf 13.36}& {\bf -0.02} & {\bf 0.50} & {\bf 0.87} \\
f &16.85 & 15.47 & 14.84 & 14.20 & 1.38  & 0.63 & 1.27\\
              &	{\bf 15.66} &{\bf 14.57} &{\bf 14.08}& {\bf 13.63} & {\bf 1.09}& {\bf 0.49} & {\bf 0.94} \\
g &	 19.14 & 17.18 & 16.58 & 15.91 & 1.96 & 0.60 & 1.27 \\          
	      & {\bf 17.95}& {\bf 16.28} &{\bf 15.82}& {\bf 15.34} & {\bf 1.67}& {\bf 0.46} & {\bf 0.94} \\
            \noalign{\smallskip}
	    \hline
         \end{tabular}
     $$
\begin{list}{}{}
\item[$^{\mathrm{a}}$] Values in bold are corrected for inclination and total Galactic foreground extinction.
\item[$^{\mathrm{b}}$] Objects are labeled according to Fig.~\ref{ident}.
\end{list}
   \end{table}

\section{Spectroscopy}

Long-slit spectra of the individual galaxies were
obtained during two different runs in September and November 2001 at the 1.82 m 
telescope of the Asiago Observatory (Italy) equipped with AFOSC, which has a spatial scale 
of 0.47 arcsec pixel$^{-1}$ over a 1k x 1k CCD. The grism n.4, chosen for its 
large spectral coverage, gave a dispersion of 4.2 \AA\ pixel$^{-1}$ in the effective 
range of 4200 -- 7700 \AA.  A slit width of 1.26 and 2.1 arcsec, selected according to the 
seeing conditions, produced a spectral resolution of $\sim$ 13 \AA\ and 24 \AA\ in the two runs, 
respectively. Typical exposure times were of 1800 sec each. See Table~\ref{log} for details 
about the observations.

Spectra were reduced with IRAF following the usual steps of bias subtraction, flat-field 
correction, cosmic-rays removal, wavelength calibration by means of Helium-Argon or 
Thorium comparison lamps and night-sky subtraction. 
Finally they were flux calibrated through the observation of the spectrophotometric 
standard stars BD+284211 and G191-b2b.
The 1D spectra obtained adding the total flux of each object with the task APALL were 
corrected for Galactic extinction using the value E(B$-$V)=0.29 derived above and applying
the \citetalias{ca89} extinction law.

Radial velocities were calculated by measuring the position of the emission-lines, where detectable,
or by using the cross-correlation technique \citet{td79} when only the absorption lines were present.
Six out of seven galaxies show comparable redshifts (Table~\ref{radvel}), suggesting
that they are members of a group; their spectra are shown in Fig.~\ref{spec}.
Only in the case of galaxy ``g'' (2MASXi~J0247387+445008) it was impossible to apply the above 
methods due to the low
S/N ratio of the spectrum. Nevertheless we could recognize some absorption lines by comparing the
spectrum with an early-type galaxy template\footnote{Galaxy template spectra by \citet{kc96}
are available from ftp://ftp.stsci.edu/cdbs/cdbs2/grid/kc96/} \citet{kc96} and we estimated a redshift 
of $\sim$ 0.17. 
This estimate is in agreement with the observed color B$-$R = 2.13, considerably redder than
typical colors for the same morphological type at low redshift \citep[see e.g. Fig.~2 of][]{l99}.
Therefore object ``g'' appears to be a background galaxy and not a member of the group.

   \begin{table}
      \caption[]{Radial velocities.}
         \label{radvel}
     $$
         \begin{tabular}{llllllll}
            \hline
            \noalign{\smallskip}
            Obj. Id. & $\alpha$ (J2000) &  $\delta$ (J2000) & V$_{hel}$  \\
	             &  ($^{\rm h~~m~~s}$) & ($^{\circ~~\prime~~\prime\prime}$) & (km sec$^{-1}$) \\
            \noalign{\smallskip}
            \hline
CG~J0247+449a & 02 47 25.2 & 44 50 39 & 10981 $\pm$ 57  \\
CG~J0247+449b & 02 47 37.6 & 44 51 45 & 11346 $\pm$ 116  \\
CG~J0247+449c & 02 47 46.3 & 44 52 58 &  11386 $\pm$ 101 \\
CG~J0247+449d& 02 47 49.8 & 44 53 26 & 12240 $\pm$ 32 \\
CG~J0247+449e & 02 47 41.9 & 44 50 34 & 11707 $\pm$ 41 \\
CG~J0247+449f& 02 47 40.4 & 44 50 29 & 11931 $\pm$ 88 \\
2MASXi J0247387+445008 & 02 47 38.7 & 44 50 08 & 51000 (?) \\          
            \noalign{\smallskip}
	    \hline
         \end{tabular}
     $$
   \end{table}

\section{Notes on individual members}

{\it CG J0247+449a.} The morphological analysis in all four photometric bands 
indicates a
regular galaxy without asymmetric structures. It is well fitted by an
exponential bulge (n $\sim$ 1) and the B/T ratio is $\sim$ 0.37. No residuals are visible
after model subtraction (see Fig.~\ref{BVRI_r}). 
For this object (also known as 2MASXi~J) we suggest a morphological type Sa/Sab.
The optical spectrum shows weak continuum and emission lines (Fig.~\ref{spec}). 
[\ion{O}{III}]5007 is
the brightest emission line, while Balmer hydrogen lines are partly absorbed by the
underlying stellar continuum. After the subtraction of this stellar
contribution by means of a template galaxy spectrum, following the
prescriptions by \citet{hfs93}, we observe an H$\alpha$/H$\beta$ ratio
$\sim$ 2.89, very close to the theoretical value of 2.86 for Case B recombination
at electronic temperature T$_e$ = 10$^4$ K \citep{ost89}, indicating very low internal
extinction. The logarithmic ratios [\ion{O}{III}]5007/H$\beta$ = 0.95$\pm$0.34 and 
[\ion{N}{II}]6583/H$\alpha$ = -0.30$\pm$0.18,
even with large uncertainties, are typical of a Seyfert-2 nucleus \citep{vo87}.

{\it CG J0247+449b.} This galaxy (2MASXi~J) shows a B/T ratio $\sim$ 0.57, almost constant in all
bands (see Table~\ref{gim2d}). The bulge is well reproduced by a de Vaucouleurs $r^{1/4}$-law, 
therefore we classify this object as E/S0.
The residuals show
in each band a ring-like structure at a radius of $\sim$ 4 arcsec 
(better visible in the R-band in Fig.~\ref{BVRI_r}) and a central 
X-structure aligned with the major and minor 
photometric axes, which indicates a disky shape of the inner isophotes \citep{schw98}.
Numerical simulations \citep{nb01} have shown that faint ellipticals with disky isophotes
might be the final product of unequal mass disk galaxy mergers (mass ratio 3:1 or 4:1).

The integrated spectrum of the galaxy shows only absorption lines and a continuum
typical of an early-type galaxy (Fig.~\ref{spec}).

{\it CG J0247+449c.} This galaxy (2MASXi~J) has a B/T ratio around or even less than $\sim$ 0.4.
Its bulge is well fitted by a de Vaucouleurs $r^{1/4}$-law like for object ``b''.
So we classify it as S0/Sa.
The residuals in VRI bands show the presence of a nuclear source, confirmed
by the analysis of the optical spectrum.
Indeed, the spectrum has the continuum of an early-type spiral (see Fig.~\ref{spec}) 
with deep metal absorption
lines but with H$\alpha$, [\ion{N}{II}]6583 and [\ion{S}{II}]6724 emissions. 
Since the [\ion{N}{II}]6583 line appears more intense than H$\alpha$ and the 
[\ion{O}{III}]5007 is barely detectable at noise level, we
suggest that this galaxy could host a LINER.
In order to avoid an underestimate of the H$\alpha$ flux due to the underlying stellar
absorption, we applied a template correction, as in the case of object ``a''.
After this correction the two emission lines H$\alpha$ and [\ion{N}{II}]6583 have
comparable intensity. The diagnostic logarithmic emission-line ratios 
[\ion{N}{II}]6583/H$\alpha$ and [\ion{S}{II}]6724/H$\alpha$ are 0.22$\pm$0.10 and 
-0.16$\pm$0.15, respectively, confirming the classification as a LINER.

{\it CG J0247+449d.} Object ``d'' is a galaxy seen almost edge-on ($i$ $\sim$ 80\degr).
Although it is generally not recommended for highly inclined galaxies, we attempted
a morphological analysis in this case, as well. 
Despite the lack of visible residuals after the subtraction of the bidimensional 
galaxy model in the four photometric bands (Fig.~\ref{BVRI_r}), the
output values of the fit (Table~\ref{gim2d}) are somehow puzzling, suggesting that the bulge-disk
decomposition mostly failed. 
Therefore we limit our comments to point out that the
galaxy is significantly asymmetric and the bulge seems to follow an exponential
law. We propose a Sbc classification for this object.
The optical spectrum is very faint, but weak extended [\ion{S}{II}]6724 and H$\alpha$ emission 
lines are measurable and their logarithmic ratio is -0.26$\pm$0.21, somewhat higher than 
expected for \ion{H}{II}-like galaxies. However the non-detection of [\ion{N}{II}]6583 and 
[\ion{O}{III}]5007 indicates a low ionization degree of the gas.
The H$\beta$ emission-line is detectable at noise level.  
We interpret this emission spectrum as an indication of a normal level of star formation
for a late-type spiral galaxy.

{\it CG J0247+449e/f.} These two galaxies form a close pair known as IRAS 02443+4437.
Despite the application of the maximum entropy algorithm to the far infrared (FIR) raw data
extracted from the IRAS database to approach the diffraction limit of the telescope,
the galaxy pair could not be resolved. A single infrared source, 
centered at the position of galaxy ``f'', is detectable only at 60 (Fig.~\ref{me}) and 100 $\mu$m
with measured fluxes 0.618$\pm$0.005 and 1.313$\pm$0.085, respectively.
The total FIR luminosity between 40 and 120 $\mu$m calculated following \citet{hsr85} is 
L$_{\rm FIR}$ = $2.75\times10^{10}$ L$_{\odot}$, which yields a total star formation rate
SFR = 14.4 M$_{\odot}$ \citep{hu86}. The spectroscopic analysis reveals that the FIR emission
is dominated by galaxy ``e'', which shows bright emission lines typical of a starburst galaxy.
Measured fluxes of the emission lines (Table~\ref{flux}) have been corrected for internal extinction
A$_{\rm V}$ = 1.61 determined assuming a theoretical H$\alpha$/H$\beta$ = 2.86 and applying 
the \citetalias{ca89} extinction law. The logarithmic diagnostic ratios [\ion{N}{II}]6583/H$\alpha$ =
-0.41$\pm$0.02,
[\ion{S}{II}]6724/H$\alpha$ = $-0.51\pm0.04$, [\ion{O}{I}]6300/H$\alpha$ = $-1.61\pm0.13$, and 
[\ion{O}{III}]5007/H$\beta$ = $-0.16\pm0.06$ confirm the thermal nature of the ionizing source.
The reddening corrected H$\alpha$ luminosity L(H$\alpha$) = $5.1\times10^{41}$ erg s$^{-1}$ corresponds
to a SFR $\simeq$ 4 M$_{\odot}$ yr$^{-1}$ \citep{kenn98}.
This value is lower than the SFR of the galaxy pair estimated from the FIR luminosity,
but it only accounts for the portion of the galaxy covered by the slit. 
The spectrum of object ``f'' shows the typical absorption features of an early-type galaxy, 
without any emission line, but with a relatively blue continuum.
The morphological analysis confirms the spectroscopic results. Galaxy ``e'' is clearly a spiral with 
irregular shape, as indicated by the high value of the asymmetry index, which made less accurate the
bidimensional fit. In fact strong residuals remain where star forming regions are located.
These regions, brighter in the nucleus and fainter in the inner spiral arms, are likely to be responsible 
for the relatively high value of the obtained B/T ratio ($\sim\, 0.37$), and for the negative B$-$V 
color (see Table~\ref{mag}).  
Since the S\'ersic index of the bulge results $<$ 1, we suggest a Sbc classification for this object.
Galaxy ``f'' shows a mean B/T ratio of $\sim\, 0.7$ and its bulge is well fitted by a de Vaucouleurs 
$r^{1/4}$-law, indicative of a morphological type E/S0. A moderate asymmetry is present
(Table~\ref{gim2d}) and the 
residuals show in all bands an arm-like structure, which departs from the south of the nucleus winding 
around it, and a blob north of the nucleus.
The morphological distortions observed in these galaxies together with their accordant redshifts indicate 
that this is a real interacting pair.
As demonstrated by some authors \citep{rs92,ht99} galaxy pairs with mixed morphology exist in significant numbers
and are believed to be the product of interaction phenomena. In agreement with our results 
about this galaxy pair, it is found 
that spiral + lenticular systems show an enhancement in the FIR emission with increased rate and 
efficiency of induced star formation \citep{ht01}.

   \begin{table}
      \caption[]{Emission line fluxes of CG~J0247+449e.}
         \label{flux}
     $$
         \begin{tabular}{lll}
            \hline
            \noalign{\smallskip}
            Line & F$_{\lambda}$/F$_{\beta}$ & I$_{\lambda}$/I$_{\beta}$   \\
            \noalign{\smallskip}
            \hline
	    $[$\ion{O}{II}$]\lambda$3727  & 5.56 $\pm$ 1.18 & 9.70 \\
	    $[$\ion{O}{III}$]\lambda$4959 & 0.30 $\pm$ 0.08 & 0.29 \\
	    $[$\ion{O}{III}$]\lambda$5007 & 0.75 $\pm$ 0.11 & 0.70 \\
	     \ion{He}{I}$\lambda$5876     & 0.21 $\pm$ 0.06 & 0.15 \\
	    $[$\ion{O}{I}$]\lambda$6300   & 0.11 $\pm$ 0.04 & 0.07 \\
	    $[$\ion{N}{II}$]\lambda$6548  & 0.50 $\pm$ 0.09 & 0.30 \\
	    H$\alpha$                     & 4.77 $\pm$ 0.36 & 2.86 \\
	    $[$\ion{N}{II}$]\lambda$6583  & 1.86 $\pm$ 0.18 & 1.11 \\
	    $[$\ion{S}{II}$]\lambda$6716  & 0.84 $\pm$ 0.12 & 0.48 \\
	    $[$\ion{S}{II}$]\lambda$6731  & 0.71 $\pm$ 0.11 & 0.41 \\
	    & & \\
	    E(B-V) = 0.52 $\pm$ 0.08 & & \\
	    I$_{\beta}$ = $6.04\times10^{-14}$ & (erg cm$^{-2}$ s$^{-1}$)  & \\
	    \noalign{\smallskip}
	    \hline
         \end{tabular}
     $$
   \end{table}

\section{Discussion}

As shown in Sect.~3 on the basis of radial velocity measurements, 6 out of the 7 galaxies 
grouped around the barycentric position $\alpha$(J2000) = $02^{\rm h}47^{\rm m}36^{\rm s}.3$, 
$\delta$(J2000) = 44\degr51\arcmin39\arcsec,
form a physical system with a median radial velocity V=11730 km s$^{-1}$.
The group, whose ellipticity calculated following \citet{rood79} is $\epsilon$ = 0.68, shows
an elongated shape, a property that have been found to be typical of galaxy groups and interpreted
as indication of three-dimensional shapes intrinsically prolate \citep[][and references therein]{hick97}.

Although the group does not exhibit a particularly compact configuration, we find that its properties
fulfill all criteria applied by Hickson \citep{hick82} to define compact groups of galaxies (CGs).
In fact the median projected separation of our group, R=121.7 kpc\footnote{throughout this paper we
assume H$_0\,=\,75~\rm km~s^{-1}$, unless otherwise specified.}, 
is significantly higher than the values observed in the most compact systems, like Seyfert's Sextet 
or HCG 31 (\cite{hick97}), but still within the range of R measured in the Hickson's sample (HCGs).
The average surface brightness in R band, $\mu_R= 24.4~\rm mag~arcsec^{-2}$, evaluated within the 
minimum circle containing the centers of all galaxy members, having radius $\theta_{\rm R} \sim$ 
2\arcmin.6, 
satisfies the compactness criterion, requiring $\mu_{\rm R} < 26.0~\rm mag~arcsec^{-2}$. 
The isolation of the group is assured by the fact that the nearest cataloged galaxy of comparable 
magnitude (2MASXi J0246441+444820) is at a distance 3.7$\theta_{\rm R}$ from the group's center, therefore 
well outside the isolation radius $\theta_N=3\theta_{\rm R}$.

Moreover, as we have shown in Sect. 4, CG~J0247+44.9 exhibits a wide range of activity types, being characterized by the 
presence of a Seyfert 2 galaxy, a LINER, a HII galaxy and a strongly interacting starburst galaxy.
The two active nuclei are hosted by early-type spirals.
This is in agreement with \citet{coz98a,coz98b}, who found in a sample of 17 HCGs that the 
AGNs are preferentially located in the most early-type galaxies and that 
50\% of them are low-luminosity ones, like Seyfert 2 or LINER. 
They also pointed out that AGNs are systematically concentrated 
toward the central parts of the groups, while starburst galaxies are distributed in the external parts.
This is not verified in our case, as the two active galaxies of CG~J0247+44.9 are located in the 
outer parts of the group, while the starburst galaxy is near the geometrical central position.
It is interesting to note that the closest object to the geometrical center of the group is  
the non-active, early-type galaxy ``b'', whose internal structures (see Sect.~4) might be the
outcome of an unequal mass, disk galaxy merger. 
 
A good agreement has also been found between other properties of CG~J0247+449 and those of HCGs.
In particular the velocity dispersion $\sigma_{\rm V}$ = 458 km s$^{-1}$ is lower than the maximum observed
value for HCGs, albeit higher than the median of 200  km s$^{-1}$ \citep{hick92}. 
The intrinsic three-dimensional velocity dispersion $\sigma_{\rm 3D}$ = 709.5 km s$^{-1}$ calculated 
taking into account the velocity errors, has been used to estimate the dimensionless crossing time
H$_0$t$_c$, which gives an indication of the dynamical state of the group.
We have found a value of 0.016 that, when compared with the fraction of late-type spirals in the 
group (f$_s$ = 0.34), results in good agreement with the $\log$H$_0$t$_c$ --  f$_s$ relationship 
found by Hickson et al. (1992). According to this relationship, the spiral fraction is lower on 
average in groups with small crossing times, as expected in case of hierarchical evolution within
galaxy groups. 
Since numerical simulations \citep{per90} have shown that a group should approach the virial 
equilibrium 
condition after about three crossing times and our group has a relatively low t$_c$, we 
calculated the virial mass following \citet{per90} and found a value of $\sim$ $1.5\times10^{13}$ 
M$_{\sun}$, comparable with the virial masses of HCGs, which reach a maximum value of $7.3\times10^{13}$ 
M$_{\sun}$.
It must be stressed that the assumption of virial equilibrium and the neglect of the mass spectrum 
of the galaxies could lead to a large overestimate of the virial mass.  
Assuming as B luminosity of the group the luminosity of all the
accordant galaxies, L$_{\rm B}$ = $1.95\times10^{11}$ L$_{\sun}$, we found a mass-to-light ratio 
M/L$_{\rm B}$ $\sim$ 77, somewhat higher than the median value for HCGs (M/L$_{\rm B}$ = 37.5), but
significantly smaller than typical values for loose groups \citep[][and references therein]{hick92}.

On the basis of the above observational properties we suggest that  CG~J0247+44.9
is a dynamically old but still evolving compact galaxy group, as indicated by the strong signs
of interaction in the close pair IRAS 02443+4437 belonging to the system. 
This idea is in agreement with the relatively high velocity dispersion of the group, typical of groups
having a high fraction of early type galaxies \citep{hick97}. These systems often show an extended 
soft X-ray halo \citep[][and references therein]{mulch00} suggesting that they are 
physically bound and -- at least in case of symmetric X-ray morphologies -- already relaxed, therefore
dynamically old groups.
The ROSAT All-Sky Survey (RASS) shows no obvious excess of X-ray emission at the location of CG~J0247+44.9.
To check whether this is consistent with the presence of a galaxy group,
 we have estimated the expected soft X-ray luminosity and temperature of our group 
 by means of the empirical L$_X$--$\sigma_{\rm V}$ and T--$\sigma_{\rm V}$ relations from 
 \citet{muza98}
 obtaining L$_X$ = $5.96\times10^{42}$ erg s$^{-1}$ and T = 1.38 keV. 
 At the distance of the source (D $\sim$ 156.4 Mpc) this luminosity corresponds to a flux of
 $\sim$ $2.03\times10^{-12}$ erg cm$^{-2}$ s$^{-1}$, a value approaching the 
 3$\sigma$ upper limit to the X-ray flux derived from RASS data within a 250 h$^{-1}$ kpc 
 (h = H$_0$/100) aperture after correction for the Galactic neutral hydrogen column density
 in the direction of the group (H. Ebeling, private communication).
 Therefore the lack of cataloged RASS X-ray sources at the location of
 CG~J0247+44.9 is not
 inconsistent with our results.

\section{Conclusions}

We have identified a new compact group of galaxies at low Galactic latitude and analyzed the 
main spectroscopic and morphological features of its members. The general properties of the 
system have been found in good agreement with those of Hickson Compact Groups. 
In summary this group, named as CG~J0247+44.9, is composed by six members, including the close 
interacting S0+S pair, IRAS~02443+4437, whose enhanced infrared emission is dominated by the spiral 
component.
Four out of the six galaxies show some kind of activity, ranging from a moderate 
level of star formation to a Seyfert nucleus. 
The closest galaxy to the geometrical center of the group is an E/S0 with internal structures
suggestive of a past history of interaction or even merger.
The early-type morphology (S0/Sa) is dominant in the 
group, which comprises only two late-type spirals. In the framework of hierarchical evolution, where 
spirals are transformed into ellipticals through interactions and mergers, this property combined with 
the short crossing time and the relatively high velocity dispersion suggests that the group is 
dynamically old. However the evident strong signs of interactions in the close pair indicate that 
the system is still evolving. High spatial and spectral resolution kinematics would be necessary
to better establish the dynamical state of this galaxy group and the interaction history of its
members.

\begin{acknowledgements}
We are grateful to Harald Ebeling for having provided us with the analysis of ROSAT data 
and for the useful discussion, and to Ronald Weinberger for a careful reading of the
manuscript and useful suggestions. 
SC is grateful to the Institut f\"ur Astrophysik Innsbruck, for warm hospitality and to the
Astrophysikalisches Institut Potsdam for the access to the IRAS data processing
with the maximum entropy program.
ST acknowledges support by the Austrian Science Fund (FWF) under project no. P15065.
We wish also to thank the technicians of Asiago Observatory for their assistance during 
the observations.
    
\end{acknowledgements}

\bibliographystyle{aa}

\clearpage

    \begin{figure}
   \centering
      \caption{R-band exposure of the galaxy system around the position 
      $\alpha$ = $02^{\rm h}47^{\rm m}36^{\rm s}.3$, $\delta$ = 44\degr51\arcmin39\arcsec (J2000). 
      The galaxies identified in the field of
      view are marked with ellipses and labeled with small-case letters. The 
      minimum circle containing all geometrical centers of the galaxies is 
      drawn, as well. Objects ``e'' and ``f'' are the components of the 
      interactive pair IRAS~02443+4437.  
              }
         \label{ident}
   \end{figure}  
   
   
   \begin{figure}
   \centering
      \caption{Thumbnails showing the BVRI-band images of the seven galaxies
      identified in Fig.~\ref{ident}. The short bars on the bottom-right corner
      of the I images of each galaxy are 5 arcsec in size.
              }
         \label{BVRI}
   \end{figure}

   
   \begin{figure}
   \centering
      \caption{Thumbnails showing the BVRI-band residuals of the seven galaxies
      identified in Fig.~\ref{ident}, after subtraction of the GIM2D bidimensional
      best-fit models. Bars are the same as in Fig.~\ref{BVRI}.
              }
         \label{BVRI_r}
   \end{figure}

\clearpage

   \begin{figure}
   \centering
   \includegraphics[width=\textwidth]{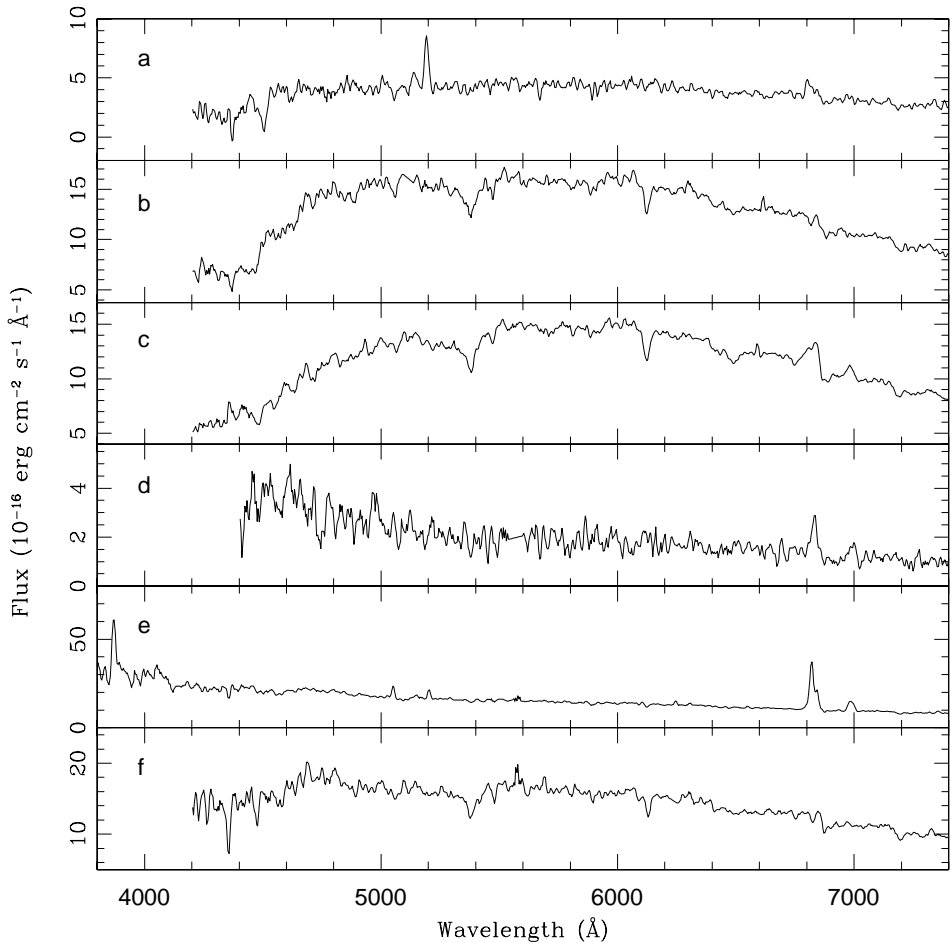}
      \caption{Optical spectra of the six galaxies identified as members of the
      galaxy group CG~J0247+44.9, labeled according to Fig.~\ref{ident}.
      Emission lines are visible in objects ``a'', ``c'', ``d'', and ``e''
      with different properties (see text for a detailed description).
              }
         \label{spec}
   \end{figure}

   
   \begin{figure}
   \centering
      \caption{Contour map of the 60 $\mu$m IRAS emission of the galaxy pair
      IRAS~02443+4437, after application of the maximum entropy algorithm,
      overplotted to the Digitized Sky Survey optical image. Dashed contours
      are drawn at 1$\sigma$ and 2$\sigma$ level over the background; solid
      contours are drawn at 3, 4, 5, 6, 10, and 20$\sigma$ levels.  
              }
         \label{me}
   \end{figure}

\end{document}